Model Inadequacy and Mistaken Inferences of Trait-Dependent Speciation

Daniel L. Rabosky[1][*] and Emma E. Goldberg[2][*]

[1] *Museum of Zoology and Department of Ecology and Evolutionary Biology, University of Michigan, Ann Arbor, MI 48109 USA*

[2] *Department of Ecology, Evolution, and Behavior, University of Minnesota, St Paul, MN 55108, USA*

[*] *The authors contributed equally to this work.*

Running title: MISTAKEN INFERENCE OF TRAIT-DEPENDENT SPECIATION

Corresponding author contact information:

Daniel L. Rabosky

Museum of Zoology and Department of Ecology and Evolutionary Biology, University of Michigan, Ann Arbor, MI 48109 USA

Email: drabosky@umich.edu

Phone: 510 - 610 - 9082

ABSTRACT

Species richness varies widely across the tree of life, and there is great interest in identifying ecological, geographic, and other factors that affect rates of species proliferation. Recent methods for explicitly modeling the relationships among character states, speciation rates, and extinction rates on phylogenetic trees- BiSSE, QuaSSE, GeoSSE, and related models - have been widely used to test hypotheses about character state-dependent diversification rates. Here, we document the disconcerting ease with which neutral traits are inferred to have statistically significant associations with speciation rate.  We first demonstrate this unfortunate effect for a known model assumption violation: shifts in speciation rate associated with a character not included in the model. We further show that for many empirical phylogenies, characters simulated in the absence of state-dependent diversification exhibit an even higher Type I error rate, indicating that the method is susceptible to additional, unknown model inadequacies. For traits that evolve slowly, the root cause appears to be a statistical framework that does not require replicated shifts in character state and diversification. However, spurious associations between character state and speciation rate arise even for traits that lack phylogenetic signal, suggesting that phylogenetic pseudoreplication alone cannot fully explain the problem. The surprising severity of this phenomenon suggests that many trait-diversification relationships reported in the literature may not be real.  More generally, we highlight the need for diagnosing and understanding the consequences of model inadequacy in phylogenetic comparative methods.

Explaining the distribution of diversity across the tree of life remains a central challenge in evolutionary biology and ecology. Some groups of organisms are spectacularly diverse, yet many other groups are species-poor. Numerous studies have demonstrated that the heterogeneity in species richness among groups of organisms cannot be explained by a homogeneous speciation-extinction process (Stanley et al. 1981; Strathmann and Slatkin 1983; Ricklefs 2003). Rather, a substantial fraction of the variation in species richness among groups appears to reflect differences in macroevolutionary rates. This general conclusion is supported by explicit modeling of evolutionary rates on phylogenetic trees, which has found considerable evidence for heterogeneous speciation-extinction dynamics among clades (Jetz et al. 2012; Rabosky et al. 2013).

Numerous studies have attempted to link differences in macroevolutionary rates to ecological, geographic, life-history, and other traits that might affect rates of speciation and extinction (Jablonski 2008; Rabosky and McCune 2010; Ng and Smith 2014). For example, clades of plants with floral nectar spurs appear to diversify at faster rates than their sister clades lacking nectar spurs (Hodges 1997). The mechanisms underlying the correlations between characters and diversification are generally poorly understood, and identifying the traits that truly influence lineage diversification requires experimental and theoretical investigations of how candidate characters affect the population dynamic and genetic processes involved in speciation and extinction. A first step, however, is using statistical methods merely to test whether such a correlation exists.  If it does, we infer that the character of interest has a functional or adaptive connection to lineage diversification, either directly or indirectly through other traits (Maddison and FitzJohn 2014).

STATISTICAL TESTS OF CHARACTER STATE-DEPENDENT DIVERSIFICATION

*Sister Clade Contrasts*

Perhaps the first formal statistical test of the relationship between a particular character state and diversification was performed by Mitter et al. (1988). In a seminal paper, they introduced sister clade contrasts as a method to test for the repeated effects of a character on diversification. The general idea is straightforward: using a phylogeny, identify a set of sister clades that differ in some key trait of interest. Each pair of clades is thus a single datum, and one tests whether the contrast in character states predicts the contrast in species richness. Mitter et al. (1988) used the approach to test whether clades of phytophagous insects contained more species than their respective sisters. In 11 of 13 contrasts, the phytophagous clade had greater diversity than the non-phytophagous clade, demonstrating a significant association between diet and diversification rate ($p = 0.01$; one-tailed sign test). This general statistical framework, with some extensions, has been widely used in comparative biology to identify correlates of diversification (Barraclough et al. 1995; Hodges and Arnold 1995; Isaac et al. 2005).

Sister clade contrasts are firmly rooted in the phylogenetic comparative method (Felsenstein 1985; Harvey and Pagel 1991). Conceptually, one can view them as a set of phylogenetically independent, paired contrasts in species richness and character state. Under the simplest model of character evolution and the simplest statistical sign test, the method of sister clade contrasts requires at least six independent shifts in character state across a phylogeny to achieve significance at the a = 0.05 level (two-tailed test). The method is intuitive and, due to the replication required, appears to be conservative. Despite this apparent statistical robustness, however, biases can be introduced by character evolution and the selection of clades separated by fixed character differences. In particular, a higher transition rate to one state can be mistaken for increased diversification associated with that state (Maddison 2006), or the waiting time before a derived state appears can cause an apparent association between that state and reduced diversification (Kafer and Mousset 2014).

*BiSSE and Related Models*

The binary-state speciation and extinction model (BiSSE) and subsequent related methods were introduced to solve a problem first identified by Maddison (2006). He noted that asymmetric rates of character state change were confounded with the effects of a character on speciation or extinction rate, making the two processes difficult to disentangle. The solution, developed by Maddison et al. (2007), was a new method - BiSSE - that simultaneously modeled character change and its effects on diversification. The BiSSE model has been extended to accommodate quantitative traits (FitzJohn 2010), geographic character states (Goldberg et al. 2011), multiple characters (FitzJohn 2012), punctuated modes of character change (Goldberg and Igic 2012; Magnuson-Ford and Otto 2012), time-dependent macroevolutionary rates (Rabosky and Glor 2010), extinct species (Stadler and Bonhoeffer 2013), and more (FitzJohn et al. 2009; FitzJohn 2012). We refer to this general family as SSE models. Numerous studies have applied SSE models to empirical datasets to identify correlates of species diversification. The methods appear to have high power for reasonably large trees, and many studies have identified significant correlations between particular character states and rates of species diversification (e.g., Lynch 2009; Goldberg et al. 2010; Johnson et al. 2011; Mayrose et al. 2011; Hugall and Stuart-Fox 2012; Price et al. 2012; Pyron and Burbrink 2014).

*Strengths and Vulnerabilities*

When transitions among character states are relatively frequent and thus few large clades are dominated by one state, BiSSE has more power than sister clade contrasts. It also makes much fuller use of the branching shape and branch lengths that comprise a phylogeny. The BiSSE approach has an important and only recently-appreciated weakness, however, highlighted by comparison with the replication required by the sister clade method. BiSSE derives its statistical power by tracking along a phylogeny the association between the trait of interest and rates of speciation and extinction, but it does not place any weight on whether independent shifts in character state are associated with shifts in diversification (Maddison and FitzJohn 2014). Consequently, a strong correlation between character states and diversification can be inferred from merely a single diversification rate shift within a phylogenetic tree, even if the shift is unrelated to the modeled character (Maddison et al. 2007, FitzJohn 2010, Maddison and FitzJohn 2014). We demonstrate below that the problem of false positives, in which a neutral trait is incorrectly inferred to be associated with diversification differences, is more insidious than generally acknowledged. It extends to trees without obvious diversification shifts and to characters that change frequently and are thus unlikely to be codistributed with causal forces. In particular, datasets exhibiting a significant association between the states of one character and diversification are likely to show the same for many other characters. These problems suggest sensitivity not only to the assumption that there are no unmodeled changes in diversification rate, but also to other violations of the model assumptions. Consequently, it appears that current models of character-associated diversification are statistically inadequate: effects that they do not include render unreliable their conclusions about the processes of interest.

Here, we first illustrate our concerns about conclusions of state-dependent diversification using a simple analysis on a real phylogeny. We then use simulated trees to highlight one possible source of trouble. Simulations on a second set of empirical phylogenies demonstrate that additional problems persist and are likely widespread. We conclude with recommendations for future empirical and methodological work.

AN EXAMPLE WITH CETACEANS

*Body Size and Whale Speciation.*

As an example of a general problem to which SSE models might be applied, we tested whether rates of species diversification are correlated with body size in extant whales. We obtained a time-calibrated tree that includes 87 species of living cetaceans from Steeman et al. (2009), and we obtained body size data for 73 of these species from Slater et al. (2010). To our knowledge, no previous study has investigated the relationship between body size and speciation in whales, although several studies have modeled speciation and extinction rates on this phylogeny without considering character evolution (Morlon et al. 2011; Rabosky 2014). To obtain a binary character for body size, we coded each whale species as "small" or "large", based on whether the mean adult length exceeded the median size across all whales (3.52 meters). The distribution of this character state across the cetacean phylogeny is shown in Fig. 1.

We fitted two BiSSE models to the cetacean data. The first model was a differential speciation model, a five parameter model constrained such that extinction rates were equal ($\mu_0 = \mu_1$) but with separate speciation rates ($\lambda_0 \neq \lambda_1$) and potentially asymmetric character transition rates ($q_{01} \neq q_{10}$). The second model additionally constrained speciation rates to be equal across character states ($\lambda_0 = \lambda_1$) but retained asymmetric character change (4 parameters). Because our candidate model set included the most complex trait-independent diversification model in the BiSSE framework, fitting the full 6-parameter BiSSE model with separate extinction rates could only lead to increased evidence for trait-dependent diversification. For clarity of demonstration, all analyses presented below use the four- and five-parameter models described above, and are thus conservative for our purposes. We fitted each of the models using the R package diversitree (FitzJohn 2012) and compared model fits using a likelihood ratio test (LRT). We corrected for incomplete sampling (FitzJohn et al. 2009) by specifying that the phylogeny included 82% of total cetacean diversity (73 of 88 species).

We found a significant effect of body size on speciation rate. The log-likelihood of the model with separate speciation rates for large- and small-bodied lineages was -255.4, versus -258.2 under a model with equal speciation rates. Given these numbers, we can reject a model with equal rates of speciation across character states (LRT: p = 0.02, df = 1).

From this analysis, we could conclude that small-bodied whales speciate more rapidly. We would not know if body size truly affects speciation rate, because our result could arise from size being merely co-distributed with a different, causal factor (Maddison et al. 2007). However, the implication would be that something about the evolution of body size has at least an indirect effect on speciation, or that body size evolves in conjunction with an alternative character that itself directly affects speciation rate.

*Simulated Characters and Whale Speciation*

Having identified a significant effect of body size on whale diversification, we now ask: might that finding reflect purely the shape of the tree, rather than the evolution of the character itself? We thus simulated neutral characters - without an influence on speciation or extinction - on the cetacean phylogeny and asked whether they correlated significantly with speciation rate. That is, we quantify the Type I error rate of the BiSSE approach on an empirical phylogeny, retaining the observed history of lineage diversification. We simulated 100 sets of binary characters on the full cetacean phylogeny under a symmetric Markov model for each of four distinct, symmetric transition rates (q). To facilitate interpretation of the transition rates, we rescaled the cetacean phylogeny such that the root divergence occurred 1.0 time units before the present. We used four values of q (q = 0.01, q = 0.1, q = 1, and q = 10), providing a gradient from rare to frequent character state changes. For the larger transition rate values, the states of the simulated trait are interdigitated on the tree (Fig. 1, triangles) and thus not obviously co-distributed with body size; codistribution is more common for the smaller transition rates (Fig. 1, squares). We retained only simulated datasets where the rare character state was found in at least 25% of tip taxa. This requirement avoids known difficulties for BiSSE when one state is rare (Davis et al. 2013). It does, however, raise issues of acquisition bias, which we discuss more fully below.

To each dataset, we fitted the five-parameter BiSSE model with state-dependent speciation and the corresponding four-parameter model with equal speciation rates as described above, assuming complete taxon sampling. As a control, we also conducted a series of simulations where a tree of identical size (N = 87 extant species, rescaled to age = 1.0) was generated under a character-independent pure-birth model, with no among-lineage variation in speciation rate and no extinction. On these pure-birth trees, we again performed 100 simulations for each value of q described above and fit the two BiSSE models. Thus, we generated two sets of results. The first analyzes neutral characters evolved on the empirical cetacean phylogeny. The second analyzes neutral characters evolved on identically sized trees simulated without among-lineage rate heterogeneity.

Of character datasets simulated on the observed cetacean phylogeny, the overwhelming majority revealed a strong association between character state and diversification, despite no such association in the simulation model (Figure 2, top row). More than 77% of the 400 character sets showed a significant ($p < 0.05$) association between character state and speciation rate, and 58% rejected the character-independent model with great confidence ($p < 0.001$). Type I error rates were somewhat lower for intermediate values of q but approached unity for both rare (q = 0.01) and frequent (q = 10) rates of character change. In contrast, Type I error rates were not appreciably elevated for datasets simulated under a pure birth process (Figure 2, bottom row): the model with $\lambda_0 = \lambda_1$ was rejected ($p < 0.05$) in exactly 5% of pure-birth simulations. Any biases present in the pure-birth trees (e.g., due to assuming a $\chi^2$ distribution with df = 1 for the significance threshold, or acquisition bias from discarding trees with a rare state) are thus minor when compared to the extreme Type I error rates observed for the cetacean tree.

While it is certainly possible that body size underlies heterogeneous speciation dynamics across whales, these simulation results clearly show that this phylogeny possesses properties such that even neutral characters, which do not influence diversification, will frequently be statistically linked to differential speciation. This is true even for rapidly evolving neutral traits, which we might have expected to be decoupled from tree shape due to numerous state transitions. We consider this more in the Discussion, but we do not have a complete explanation. We further illustrate below that many real datasets share this unfortunate property of high Type I error for rapidly-evolving neutral traits. First, though, we use simulated trees to demonstrate one possible cause of trouble for more slowly-evolving traits: unaccounted-for heterogeneity in the diversification process. A propensity for "false positives" from analyses with BiSSE-like models is of broad concern for attempts to use phylogenetic comparative data to assess the prevalence of species selection and the traits consistently tied to speciation or extinction. We conclude with a discussion of possible ways to diagnose and address this problem.

UNACCOUNTED-FOR SPECIATION RATE HETEROGENEITY, AND
SIGNIFICANCE WITHOUT REPLICATION

There are many ways in which empirical datasets may reflect dynamics more complex than a constant-rates birth-death process, even one in which a character affects diversification. We focus here on one possible violation of the BiSSE model assumptions: shifts in diversification dynamics that may be unrelated to the character being analyzed. This case in particular is useful for understanding some strengths and weaknesses of the BiSSE approach.

Consider a phylogeny in which a high speciation rate "foreground" clade is nested within a slowly speciating "background" clade. An idealized example is shown in Fig. 3A. An empirical example is provided by cetaceans, where there is strong evidence for an increase in diversification rates somewhere along the lineage leading to the dolphin subclade (Fig. 1) (Rabosky 2014). We also generated similar phylogenies by simulation (Fig. 3BC), setting a lower speciation rate for the ancestral state ($\lambda_0 = 0.5$), a higher rate for the other state ($\lambda_1 = 1$), excluding extinction ($\mu_0 = \mu_1 = 0$), and making character transitions symmetric and either extremely rare ($q_{01} = q_{10} = 0.001$, Fig 3B) or somewhat more common ($q_{01} = q_{10} = 0.1$, Fig 3C). We required each simulated tree to have 200 tips and at least 25% representation of each state at the end time, to ensure that the tree shape contained substantial diversification rate heterogeneity. We then rescaled each simulated tree to age 1.0 and evolved neutral characters, as before, at a range of low and high rates ($q = 0.01, 0.1, 1, 10$). We kept only realizations with at least 10% of tips in each state; these are large trees, so we could use a lower threshold than in the cetacean example to reduce concerns about acquisition bias (discussed below) without making one state extremely rare. We again conducted likelihood ratio tests comparing a 5-parameter model with state-specific speciation rates (with equal extinction rates and unequal transition rates) to a 4-parameter model with equal speciation rates. Results are shown in Fig. 4.

Any character states that happen to differ in frequency between the foreground and background groups will potentially correlate with rates of speciation. In the extreme case, a single shift in diversification dynamics and a single, but independent, transition of the character (squares in Fig 3A; approximated by "rare shifts" with low q in Fig 3B and Fig 4) can generate a statistically significant association between the character and speciation rate. This result echoes previous cautions (Maddison et al. 2007; FitzJohn 2010; Maddison and FitzJohn 2014). As the diversification shifts and character transitions become more frequent, the propensity for false positives declines because chance plays a greater role in decoupling the trait from the diversification history ("common shifts" and/or high q in Fig 3 and Fig 4). In a control set of simulations, with the same speciation rate for foreground and background clades, the Type I error rate is low as expected ("no shifts" in Fig 4).

We also find that BiSSE's power to infer state-dependent diversification for the trait that truly influences speciation is no greater when the character---and consequently speciation rate---changes more frequently. The method performs very well when its assumptions are met and the signal is strong ("real trait" in Fig 4), whether shifts are rare or common ($p < 0.05$ in 99% and 97% of simulations, respectively). In a separate set of simulations for which the signal is less strong, BiSSE more frequently fails to identify state-dependent speciation for the real trait, as expected, but the reduction in power is surprisingly somewhat greater when shifts are common. For example, when $\lambda_0$ is increased to 0.75 and $\lambda_1$ remains at 1, state-dependent diversification is identified with p < 0.05 in 77% of simulations with rare shifts but 42% with common shifts.  We thus see that BiSSE does not derive its power from the association between states and speciation rates arising repeatedly; power instead comes from the total amount of tree along which the association occurs.

That BiSSE and other correlative tests for discrete characters do not require phylogenetically independent events for statistical significance has recently been highlighted as a serious concern (Maddison and FitzJohn 2014).  Focusing on Pagel's (1994) test of correlated character evolution, Maddison and FitzJohn use intuitive examples with low transition rates to show that "within-clade pseudoreplication" greatly elevates Type I error.  Our simulation results with low transition rates (small q and rare shifts in Fig. 4) show the same unfortunate effect: a chance correlation between speciation rate and a neutral trait persists through phylogenetic inertia and is judged statistically significant.  This illustrates one form of model inadequacy: speciation rate heterogeneity tied to a character not under study can drive mistaken inference of state-dependent diversification (also demonstrated by FitzJohn 2012).  Our results for simulated trees with high transition rates (large q and common shifts in Fig. 4) show reduced risk of false positives, as expected due to lesser phylogenetic inertia. Importantly, however, our simulations of neutral traits on the cetacean tree reveal even greater Type I error rates that do not systematically diminish for higher transition rates (Fig. 2).  We thus suspect that empirical phylogenies carry additional violations of the model assumptions, leading to even greater unreliability in inference of state-dependent diversification.

A POTENTIALLY WIDESPREAD PROBLEM

To assess the extent to which traits may be erroneously linked with diversification on real phylogenetic trees, we performed two additional exercises.  In the first exercise, we simulated the evolution of neutral character states on subtrees drawn from large time-calibrated phylogenies for four major vertebrate clades: birds (Jetz et al., 2012; 6670 species), ray-finned fishes (Rabosky et al. 2013; 7428 species), amphibians (Pyron and Wiens 2013; 3351 species), and squamate reptiles (Pyron and Burbrink 2014; 4451 species). For the bird tree (Jetz et al. 2012), we used the Hackett backbone phylogeny (Hackett et al. 2008) and excluded all species for which no genetic data were available, leaving a time-calibrated phylogeny of 6670 species whose positions have been estimated using at least some genetic information. We accounted for incomplete taxon sampling in our analyses using approximate diversity totals for each major group (sampling fractions of 0.667 for birds, 0.27 for fishes, 0.48 for amphibians, and 0.44 for squamates), largely following species totals presented in the references above. We partitioned the four phylogenies into all rooted subtrees that contained between 200 and 500 tips, resulting in a total of 60 bird subtrees, 36 squamate subtrees, 61 fish subtrees, and 29 amphibian subtrees.  The subtrees for each major group do not comprise a statistically independent set, because some clades are present in multiple subtrees.

We simulated binary traits on this set of phylogenies using a symmetric Markov model with four transition rates (q = 0.01, 0.1, 1.0, and 10.0), after rescaling each subtree to a root age of 1.0. We simulated 10 character histories for each combination of subtree and transition rate, giving a total of 7440 simulated datasets. We required that the rare character state occur in at least 10% of taxa for a given simulation to be accepted. Each simulation was thus conducted on a fixed topology and the simulation model specified no effect of character state on diversification. Each dataset was then analyzed using the four- and five-parameter BiSSE models described previously.

In the second exercise, we analyzed the effect of a purely arbitrary character on speciation rates across the 200 to 500 taxon subtrees drawn from the four vertebrate clades described above. We tested whether taxon name length - the number of letters in the Latin binominal for each taxon - was associated with speciation rate. We counted the number of letters in each taxon name and scored each species as "short" or "long" depending on whether the taxon name length was less than or greater than the median name length for taxa in each subtree. This character exhibits some phylogenetic signal as would an evolving trait, owing to the correlation in name lengths between congeners. For example, within the 60 bird subtrees, we found that 44 trees (73%) showed significant ($p < 0.05$) phylogenetic signal in taxon name length, as assessed by computing the K-statistic (Blomberg et al. 2003) for each dataset and determining significance via tip randomization. Name length of course cannot plausibly be considered a driver of speciation, although species richness could be reflected in linguistic or taxonomic practices. We fitted the four- and five-parameter BiSSE models described above to each subtree.

In the first exercise, we observed a high frequency of association between neutral characters and speciation rate (Fig. 5). Pooling results across taxonomic groups, we found that 61.5% of all simulated subtree/character state combinations showed a significant effect of the neutral character on speciation ($p < 0.05$). Error rates differ among the four transition rates but are surprisingly greatest when the neutral character evolves rapidly (Fig. 5). A substantial proportion of datasets were found to have highly significant ($p < 0.001$) trait-dependent diversification (20 - 32% for q = 0.01, 0.1, and 1; 73.7% for q = 10). High Type I error rates are observed across a range of character state frequencies (Table 1), indicating that this phenomenon is not driven by acquisition bias associated with requiring the character states to have similar frequencies.

In the second exercise, we found a strong effect of taxon name length on speciation rate for a majority of phylogenetic trees in each of the four major groups of vertebrates considered (Fig. 6). Results are roughly comparable to those for binary characters simulated under an explicit trait evolution model (Fig. 5). For all groups, more than 69% of trees showed a significant ($p < 0.05$) correlation between taxon name length and speciation rate; for fishes, this approached 100% (60 of 61 subtrees). The overall trend is clear: within subtrees of the four major groups of vertebrates, even arbitrary characters often exhibit a significant statistical association with speciation rates. We thus see that empirical phylogenies are even more prone than simulated ones to mistaken conclusions of state-dependent diversification.

WHAT IF TRAITS LACK PHYLOGENETIC SIGNAL?

If pseudoreplication and codistribution drive the spurious relationship between character states and speciation (our results above, and Maddison and FitzJohn 2014), we should expect this effect to be reduced or eliminated when traits evolve rapidly. We do indeed find that false positives diminish greatly for high transition rates on trees simulated with a controlled amount of speciation heterogeneity ($q = 10$ in Fig. 4). In contrast, however, the Type I error rates are very high for real phylogenies with fast evolution of a neutral character ($q = 10$ in Fig. 2 and 5). To test the limits of this surprising result, we assessed whether a purely random character---completely lacking phylogenetic signal---could be incorrectly associated with speciation rate differences.

On the whale phylogeny, we assigned tip states randomly for rare-state frequencies of 0.1, 0.2, 0.3, 0.4, and 0.5, replicating each randomization 200 times. We performed the same exercise on simulated pure-birth trees of the same size as the whale phylogeny (cf. Fig 2). For each simulated dataset, we fitted the four- and five-parameter BiSSE models described above. On a technical note, we found that it was important to perform multiple optimizations with widely-varying starting parameters when fitting the five parameter BiSSE model to trait data lacking phylogenetic signal, owing to the presence of multiple optima on the likelihood surface.

We found that every permutation of tip states for all frequencies (1000 in all) on the whale phylogeny led to a significant association between the trait and speciation rate, with 99.6% of simulations significant at the $p < 0.001$ level (Fig. 7). In contrast, the trees simulated under a pure-birth process did not show elevated Type I error rates. The false positives for the randomized traits are in accord with those for rapid transitions ($q = 10$ in Fig. 2), and they are unlikely to be due to the chance association between character states and any particular clade on the whale tree.

We also repeated the analysis described above for the whale tree, but setting just a single species to have the rare state (i.e., 86 taxa in state 0, 1 in state 1). A corresponding set of control simulations were conducted on 200 phylogenies simulated under a pure-birth process, with the rare state assigned to a randomly-chosen tip. On both the whale and the pure-birth trees, the speciation rates estimated for the two states under the five-parameter model often differed greatly (cf. Davis et al. 2013). Only on the empirical tree, however, was the fit of the four-parameter model significantly worse. Trait-dependent speciation was significantly favored on all 87 possible trait distributions on the whale tree ($p < 0.0001$) but only on 1.5% ($n = 3/200$; $p < 0.05$) of pure-birth trees.

We suspect that insufficiently accounting for phylogenetic psuedoreplication is a major component of the BiSSE method's vulnerability to false positives. It is not, however, clear how the intuitive explanation provided by Maddison and FitzJohn (2014) applies when phylogenetic signal is low or absent. We discuss below some possibilities, but we do not have a complete explanation.

DISCUSSION AND RECOMMENDATIONS

SSE models have revolutionized phylogenetic comparative tests of state-dependent diversification and the tests of character evolution with which they are entangled. The SSE method is model-based and thus provides for formal statistical parameter estimation and hypothesis testing. The framework also allows incorporation of alternative descriptions of the processes and character values, and the original BiSSE model (Maddison et al. 2007) has inspired several extensions (FitzJohn et al. 2009; FitzJohn 2010; Rabosky and Glor 2010; Goldberg et al. 2011; Magnuson-Ford and Otto 2012; Goldberg and Igic 2012; FitzJohn 2012; Stadler and Bonhoeffer 2013). These models have been shown to perform very well on simulated datasets that are reasonably large and follow the model assumptions (Davis et al. 2013; and model references above), and they have been employed in hundreds of empirical studies.

The results presented here, however, indicate that statistical SSE-based tests about the relationship between character states and speciation should be performed and interpreted with much more caution than is commonly employed. The caveat that a statistical association between a trait and lineage diversification is not evidence of a causal connection is of course as old as the statistical methods themselves (Mitter et al. 1988, p. 114; Maddison et al. 2007, p. 708). We have shown, however, that a significant association between neutral or arbitrary characters and speciation can arise with disturbing ease (Fig 2, 4-7), casting doubt on the utility of this approach to uncover traits with likely biological connections to speciation or extinction. This extends even to neutral traits that evolve rapidly and are thus expected to be decoupled from factors that truly control diversification, especially on real phylogenies, which are more likely to deviate from model assumptions than are simulated trees.

Our results are not specific to likelihood ratio tests: in the Appendix (available through Dryad; doi:10.5061/dryad.kp854), we report similar results using AIC-based model selection and using Bayesian inference of the difference in state-specific speciation rates. Although we report results only for model comparisons with BiSSE, we expect that such undesirable associations between character states and diversification rates arise with the other SSE models. Similarly, we report results only for tests of speciation rate differences, but we expect the effects documented here to extend to extinction rates, especially considering they have already been shown to be sensitive to model mis-specification (Rabosky 2010). Consequences for parameter estimation and ancestral state reconstruction are beyond the scope of the present article, but we also expect them to be substantial because when one aspect of a model is misled, other components may be warped to compensate.

Like Maddison and FitzJohn (2014), we believe that the propensity for Type I error identified here is sufficient to warrant serious and continuing discussion of the root causes and the extent to which they can be repaired. We highlight here suggestions for more robust analyses and future methodological research. This includes expanding the processes present in phylogenetic models, requiring additional diagnostic tests and replication across clades, and further statistical study of the phylogenetic comparative data structure.

*Sources of Trouble*

*Diversification rate heterogeneity*. — Diversification rate variation is ubiquitous, and numerous studies have documented highly heterogeneous speciation and extinction dynamics among lineages within phylogenetic trees (Alfaro et al. 2009; Jetz et al. 2012; Rabosky et al. 2013). In fitting a basic SSE model, however, all heterogeneity in diversification rate on the phylogeny can be attributed only to the characters included in the analysis. It has previously been recognized that a spurious finding of trait-dependent diversification can be caused by a diversification shift unrelated to the focal trait, whether due to extrinsic forces or tied to an unmodeled character (FitzJohn 2010; FitzJohn 2012). Our simulations (Fig. 3, 4) demonstrate that this problem is of a magnitude that is unappreciated in the current literature. Our results for high transition rates and randomized tip states on real phylogenies (Fig. 2, 5, 7, in contrast to Fig. 4) further highlight the statistical willingness of the BiSSE model to incorrectly assign deviations from a simple multitype birth-death process to traits that carry no phylogenetic signal. We do not have a full explanation for this behavior, but we suspect that when transition rates are high, the distribution of character states across the internal structure of the tree is largely unconstrained by the tip distribution. This allows the model to assign a "fast speciation" character state to portions of the tree where speciation rates are fast, and a "slow speciation" state to portions of the tree where speciation rates are slow. Hence, the model effectively becomes one where the characters themselves are irrelevant, and character state probabilities across internal nodes and branches are driven by diversification rate variation rather than the distribution of traits at the tips of the tree.

One possible solution might entail using a partitioned SSE model to account for some of the diversification rate heterogeneity in a phylogeny. FitzJohn (2010) used MEDUSA (Alfaro et al. 2009) to identify a diversification shift on a primate phylogeny during an analysis of the relationship between speciation rate and body size. He then assigned separate QuaSSE models to two regions of the tree that were found to have significant differences in diversification, thereby removing one violation of the QuaSSE model assumptions. In principle, one could use BAMM (Rabosky et al. 2013; Rabosky 2014) or MEDUSA to identify diversification rate heterogeneity for subsequent partitioning within an SSE model.

We caution, however, that a lack of evidence for diversification rate variation within a phylogenetic tree using BAMM or MEDUSA does not provide strong evidence that the dataset in question meets the assumptions of an SSE model. Rather, it simply means that the specific statistical models implemented in those approaches are unable to identify distinct partitions of the phylogeny that are characterized by heterogeneous diversification dynamics. As an example, we used BAMM to model diversification rate heterogeneity across the Jetz et al. (2012) avian phylogeny (see Appendix), to test whether inflated error rates persist even on subtrees with weak or no evidence for clade-specific diversification dynamics. This is the set of subtrees for which BAMM or MEDUSA would be unlikely to yield support for partitioned SSE analyses. Within this small set of phylogenies, 25 - 55% of simulations found a significant association between character states and diversification rates (Table 2). It appears that partitioned analyses have some potential to reduce error rates from SSE analyses, but we conclude that they are not a complete solution. Not only are there difficulties in fitting the more complex models, but Type I errors are found even in datasets which are not identified as problematic by these methods. Moreover, the severe pathologies that appear when character states are rare or lack phylogenetic signal may not be addressed by this approach.

When diversification rate heterogeneity is due to attributes of species, rather than for example extrinsic geologic events, including the causal character in the analysis allows the SSE model to assign diversification signal to it rather than to a focal but perhaps neutral trait (FitzJohn 2012). The difficulty, of course, is knowing which characters to consider, and the danger is that the more that are included---even if they are neutral---the more likely is a diversification effect to be attributed to one of them; FitzJohn (2012) demonstrated this with one example. Our simulations show high Type I error in many more situations even for single neutral traits, and the probability of a false positive within a collection of independent neutral traits will naturally be higher. Thus, although expanding analyses to additional traits can identify those most closely correlated with diversification differences, substantial difficulties remain for identifying the ones worthy of further study as potentially causal factors.

Alternatively, one could include an additional but uninformative trait as a stand-in

for the unknown causal force (Maddison and FitzJohn 2014). This approach is known as

the covarion model in the context of DNA sequence evolution (Fitch and Markowitz

1970) or the hidden rates model in the context of binary characters (Beaulieu et al. 2013).

The hidden trait is assigned an unknown state at all tips, and its inclusion in the model creates an additional set of rate classes. With BiSSE, diversification rate heterogeneity could thus be assigned to the uninformative trait rather than the trait of interest, potentially reducing the chances for false positives of the focal trait. In preliminary tests, we found that this hidden character procedure can indeed divert signal from a focal neutral character. It is only sufficiently effective, however, when the diversification rate heterogeneity is structured like an evolving trait (i.e., on our simulated rate shift trees;

results not shown). When that is not the case (e.g., on the empirical trees), shifts in diversification rate remain inadequately modeled and problematic for BiSSE.

*Further model assumption violations*. — Comparing our results from simulated and empirical trees raises the further concern that inadequacies in the model beyond speciation rate shifts could cause similar problems but have yet to be diagnosed. For neutral traits simulated with high transition rates or randomly arranged on real phylogenies, we observed unexpectedly high Type I error rates (Fig. 2, 5, 7), in contrast to results on simulated trees where the rapidly-evolving neutral trait is decoupled from the causal trait (Fig. 4).

The results on empirical trees are counterintuitive, because we expected that the Type I error problem would be reduced with increasing character lability across the tree. In the whale example, for instance, we might have expected greatest Type I error when the neutral character varies in frequency between the two parts of the tree with statistically distinct diversification dynamics, the dolphins and non-dolphins (Fig. 1; Rabosky 2014). Surprisingly, we find elevated Type I error rates in simulated datasets where character frequencies are virtually identical between dolphins and non-dolphins. Of the 100 simulations underlying the results in Figure 2 for q = 10, 20 had roughly equal frequencies (rare state frequency > 0.45) in the "dolphin" and "non-dolphin" taxa. All 20 simulated datasets found significant evidence for trait-dependent diversification (p < 0.05). The most extreme Type I error rates we have observed occur in the absence of any phylogenetic signal (Fig. 7). This effect does not appear with high transition rate simulations (q = 10) for our simulated datasets (Fig. 3-4), suggesting that real phylogenies reflect additional unmodeled processes, beyond the discrete speciation rate shifts we focus on explicitly. We cannot explain this effect at present, but one possibility is the presence of temporal variation in speciation rates. A recent study using the QuaSSE method reported elevated Type I error rates when phylogenies were simulated under a model of declining speciation rates through time (Machac 2014), and we note that there is substantial evidence for a slowing of speciation through time within the dolphin clade (Rabosky 2014).

In addition to temporal rate heterogeneity, any number of additional processes might warp the shapes of trees away from birth-death expectations. These include complex interactions between multiple traits and speciation, species interactions, and historical events that influence diversification dynamics. Similarly, trait evolution dynamics that are more complex than Markovian or diffusion processes could yield misleading conclusions when fit with existing simple models. Possible examples include discrete characters that evolve under a model where the probability of state change reflects the evolution of an underlying latent continuous variable (threshold models; Felsenstein 2012; Revell 2012), traits that undergo deterministic increases and decreases, or character transition rates that are dependent on previous character states and thus violate the memoryless property. Factors associated with tree construction, divergence time estimation, and taxon sampling (Pybus and Harvey 2000) can also affect the shapes of phylogenetic trees and potentially lead to to spurious trait-diversification associations.

These are all speculations, but we note that it has previously been shown that some violations of SSE model assumptions render estimates of extinction rates (Rabosky 2010) and speciation rates (Machac 2014) and tests of irreversibility (Goldberg and Igic 2012) unreliable. The situation is analogous to the problem that BiSSE was originally introduced to solve, where failing to account for one process (either directional character evolution or state-dependent diversification) misled estimation of the other (Maddison 2006; Goldberg and Igic 2008).

These cautions are diffuse because it is entirely possible that many of the assumptions to which SSE models are sensitive remain unidentified. We simply cannot know how sturdy its conclusions are to various complications until we test them. This general concern extends equally to other frameworks beyond SSE. New macroevolutionary and phylogenetic models are typically tested for power and bias using simple simulation scenarios, but rarely are new models tested for robustness to potentially complex violations of their assumptions. It would be extremely valuable for both the developers and users of all phylogenetic comparative methods to test new and old methods against a more comprehensive compilation of potential problems.

We also stress that the substantial difficulty in accurately inferring the action of character state-dependent diversification is not a justification for ignoring it. In addition to the enormous biological interest in identifying traits that affect rates of speciation or extinction, failing to account for such effects can lead to incorrect conclusions about character evolution (Maddison 2006; Goldberg and Igic 2008) and potentially other macroevolutionary processes. More robust techniques are thus necessary for both biological and methodological reasons.

*Improved Procedures*

Considerable effort will be required to improve the use of SSE models in the inference of trait-dependent speciation and extinction rates. The directions outlined above revolve around expanding the set of processes incorporated in the model being fit. A different approach is to improve the procedures used to fit the existing, simpler models and diagnose their behavior in a given dataset.

*Simulate characters*. — For any dataset at hand, it is straightforward to simulate on the phylogeny the evolution of characters that do not influence diversification. Statistical tests like we conducted here then immediately reveal whether the shape of the phylogeny itself makes it prone to Type I errors. This procedure could potentially be employed to adjust the significance threshold for tests involving the focal trait. That is, one could simulate neutral traits on a phylogeny of interest and use the distribution of likelihood ratios from their model fits to estimate the critical value for the desired level of significance. The initial tests of BiSSE's performance employed an analogous procedure (Maddison et al. 2007, p. 706). This effectively shifts the criterion for significance from absolute to relative terms: the goal becomes to identify characters that are more associated with diversification than typical, given a particular phylogeny. Although this approach warrants further investigation, we caution that even in the absence of the appreciable Type I error shown here, parameter estimates may be greatly compromised by model inadequacy, and additional diagnostics may well be required. Furthermore, it is not clear which models or transition rate values are appropriate for simulating such traits on trees (e.g., whether one should allow for state-dependent diversification or character transition rate heterogeneity when estimating the transition rates and performing the neutral character simulations).

*Multi-clade meta-analysis*. — One possibility for robust inference using SSE models is to apply the model to the same trait in multiple independent clades. These results can then be combined in a meta-analytic framework. If associations between the character and diversification are statistical artifacts, there is generally no reason to expect that the direction of the association---for example, large trait values with more rapid speciation---would be consistent across groups. Like sister clade comparisons, this approach has the advantage that replication is explicit and required.

Mayrose et al. (2011) used the meta-analytic approach to test the effects of polyploidization on diversification in 63 genera of vascular plants. By testing significance on the overall distribution of results from separate SSE analyses on each clade (specifically, a t-test on the proportion of MCMC samples for which one rate was greater than the other), they showed that polyploids consistently had lower rates of diversification than diploids. Other studies have applied SSE models to multiple sets of clades (Rolland et al. 2014) and these results could similarly be combined into a formal statistical test. Such meta-analyses could be used to detect consistent directional effects of characters on diversification across multiple datasets.

Although the phylogenetic meta-analysis approach requires extensive data, it is potentially a straightforward means around the problematic associations that we identified here. In our tests on simulated trees, on realizations that showed a significant association between the neutral trait and speciation, the direction of that association was not significantly consistent across realizations when the transition rate was moderate or high (for the $q = 1$ and $q = 10$ results in Fig. 4, $p > 0.05$ for an exact binomial test). However, when the transition rate was lower, on the trees for which the unequal speciation rate model was preferred, most showed a lower speciation rate for the root state ($p < 0.001$ for the $q = 0.01$ and $q = 0.1$ results in Fig. 4). This was true for the trees simulated with rare diversification shifts, common shifts, and no shifts. We attribute the effect to acquisition bias, described next. Furthermore, we found that on the empirical trees with unequal state frequencies (Fig. 7), the rarer state was generally associated with the higher speciation rate. We conclude that care should be taken with the meta-analysis approach when tip state frequencies differ greatly, and when clades are expected to have a consistent root state of the focal character and are chosen to have sufficient representation of the derived state. This latter bias is likely to be most pronounced when the meta-analysis framework reveals a consistent association between high diversification and the derived character state. We note that ploidy and breeding system in plants are two examples of characters thought to influence diversification and exhibit a consistent basal state across polymorphic clades (Mayrose et al. 2011; Igic et al. 2008; Goldberg et al. 2010), but for these traits it appears that character transitions are frequent and the derived state is inferred to have a lower net diversification rate, alleviating this concern.

*Dealing with acquisition bias*. — The very process of selecting clades and traits for analysis has the potential to affect statistical conclusions. In our simulations, we kept and analyzed only realizations in which both character states were present with sufficient frequency among the tip states. This reflects both statistical necessity (e.g., BiSSE may perform badly when one state is very rare (Davis et al. 2013)) and empirical practice. Our "control" simulations (pure-birth trees in Fig. 2 and trees with no shifts in Fig. 4) indicated that our own acquisition bias did not drive the magnitude of Type I error we reported, although we acknowledge that the bias may have different effects on trees with different properties.

Within the trees showing false positives, however, acquisition bias can explain the consistent association of the root state with the lower speciation rate, described in the multi-clade analyses above. The non-root state will only attain noticeably high frequency in realizations where it happens to arise in a lineage that eventually diversifies more than average. This effect persists when the foreground clade has a lower, rather than higher, speciation rate, but its magnitude is reduced (results not shown).

We can reason through some such consequences of acquisition bias, but a general solution is lacking. Lewis (2001) proposed a statistical correction for the special case of requiring at least one tip in each state, but a more general procedure for modeling what draws biologists to a particular comparative dataset is entirely unclear. This problem extends beyond tests of state-dependent diversification (e.g., Goldberg and Igic 2008), but although it is a widespread issue, it is likely only one cause of trouble among many. Our findings of incorrect trait-diversification associations, especially those that involve randomizations of various tip frequencies (Fig. 7), cannot be fully explained by acquisition bias.

*Broader Concerns*

All models make assumptions that are violated by real-world data, but the concern here is that the answers we are especially interested in obtaining from SSE models are not robust to some such violations. This leads to two immediate questions, which apply to phylogenetic comparative methods beyond the SSE framework. How do we know when to be suspicious of particular findings, so that we can search for additional factors or processes that should be incorporated in the analysis? What is the root cause of this apparent propensity for spurious results, and how can we avoid it?

The first question is often addressed on a case-by-case basis with empirical studies. When different methods yield different biological conclusions from the same dataset or when the answers conflict with biological expectations (Takebayashi and Morrell 2001, Igic et al. 2006, Syme and Oakley 2012, Miglietta and Cunningham 2012), one might become suspicious of the methods and investigate them further. More generally, however, better tools are needed for testing the goodness-of-fit of phylogenetic and macroevolutionary models. Instead of choosing the best among a set of models, no matter how insufficient they all are, one could identify situations where there is no existing adequate model. This is a basic component of the standard statistical toolbox that remains largely absent from the phylogenetic world, although posterior predictive approaches are now providing a way forward (Slater and Pennell 2014, Pennell et al. unpub.).

The answer to the second question may lie in the issue we discussed above: SSE models do not require multiple independent changes in character state or diversification in order to detect a significant effect (Maddison and FitzJohn 2014). There is simply a likelihood under a particular model, which includes no accounting for the number of independent shifts in character state and diversification  Put another way, some phylogenetic models for detecting correlated changes across phylogenetic trees---including ones much older than SSE (e.g., Pagel 1994)---do not have a clear definition for or means to assess the effective sample size or degrees of freedom.  Sample size is probably not defined solely by the tree itself (e.g., number of tips or total branch length) but instead depends also on the character being studied and its distribution across the clade. This is a serious statistical concern that lacks clear resolution. This concern does not apply to the use of independent contrasts (Felsenstein 1985) for detecting correlated character changes, which explicitly accounts for independent changes in character across the phylogeny via the calculation of contrasts at each interior node of the tree. However, we note that even the non-independence issue is unlikely to explain our results in full: our finding that Type I error rates are greatly elevated even when characters lack any phylogenetic signal suggests that other statistical issues remain to be identified.

The sample size issue is exactly the problem that is circumvented by the many-clade meta-analysis approach discussed above. Other approaches, pre-dating the SSE framework, also explicitly require replication. This includes sister-clade contrasts and tests that reconstruct separately the locations on the tree of character state changes and diversification shifts (Ree 2005; Moore and Donoghue 2009). These methods have their own inherent problems, notably how sister clades are chosen (Maddison 2006; Kafer and Mousset 2014) and how ancestral states are reconstructed. Renewed attention to such approaches may be fruitful, however, with an eye to further testing them against their own assumptions and to developing hybrid approaches (e.g., using ancestral state reconstruction from SSE models in Ree's [2005] method). It is quite possible, however, than an entirely new approach is needed for robust tests of whether a character affects rates of speciation or extinction.

CONCLUSIONS

These results call into question a large body of literature that has documented associations between character states and diversification rates. The point of the present paper is not to review these past studies or to point out specific instances of problematic results. The issues raised here are likely to be important in the interpretation of many published studies that have used SSE models to infer the relationship between character states and diversification. Until a more satisfactory solution is found, we recommend that analyses that include SSE models use a multi-clade framework for the greatest robustness, or at least explicitly address the propensity for Type I error for a given phylogeny using neutral trait simulations similar to those performed here. Despite the methodological difficulties, however, identifying state-dependent diversification remains important in its own right and for recognizing its confounding effects in studies of character evolution.  Most generally, we call for much greater attention to the diagnosis and consequences of model inadequacy in phylogenetic comparative methods. The results of comparative analyses are generally non-intuitive and lack the standard battery of visual and numerical diagnostics that are applied to linear regression and other traditional statistical analyses. As such, researchers who utilize phylogenetic comparative methods must be acutely aware of the potentially serious consequences of model inadequacy in real datasets.

## SUPPLEMENTARY MATERIAL

Supplementary material, including R code for simulation and analysis, can be found in the Dryad data repository (doi:10.5061/dryad.kp854).

## ACKNOWLEDGEMENTS

We thank T. Barraclough, Y. Brandvain, S. Heard, B. Igic, S. Otto, V. Savolainen, and two anonymous reviewers for extensive comments on earlier versions of this manuscript, and W. Maddison and R.  FitzJohn for enlightening discussions.  This work was carried out in part using computing resources at the University of Minnesota Supercomputing Institute. This work was supported in part by NSF DEB-1256330 to D. L. R.

Table 1. Type I error rates for neutral character simulations on 186 empirical phylogenies of birds, squamates, amphibians, and fishes. Results are binned by frequency of the rarer character state and are pooled across transition rates and clades. Data are identical to those presented in Figure 5. The second column gives the number of simulated realizations. The third and fourth columns report the proportion of realizations for which state-independent diversification is rejected.

| Character freq. | N | p < 0.05 | p < 0.001 |
|---|---|---|---|
| $0.10 \le x < 0.15$ | 1562 | 0.404 | 0.164 |
| $0.15 \le x < 0.20$ | 713 | 0.456 | 0.212 |
| $0.20 \le x < 0.25$ | 444 | 0.617 | 0.387 |
| $0.25 \le x < 0.30$ | 241 | 0.631 | 0.402 |
| $0.30 \le x < 0.35$ | 291 | 0.632 | 0.423 |
| $0.35 \le x < 0.40$ | 171 | 0.608 | 0.298 |
| $0.40 \le x < 0.45$ | 163 | 0.669 | 0.423 |
| $0.45 \le x < 0.50$ | 128 | 0.656 | 0.414 |

Table 2. Type I error rates (p < 0.05) for BiSSE analyses of the avian subtree dataset for subtrees where BAMM was unable to identify a strong signal of diversification rate variation.

| Transition rate | BAMM $p_0$ > 0.05 (N = 70) | BAMM $p_0$ > 0.25 (N = 40) |
|---|---|---|
| q = 0.01 | 0.300 | 0.275 |
| q = 0.1 | 0.257 | 0.300 |
| q = 1 | 0.343 | 0.325 |
| q = 10 | 0.542 | 0.500 |

Notes: BAMM $p_0$ is, for each avian subtree, the posterior probability of a model with zero diversification rate shifts as inferred using BAMM (Appendix). Hence, the column "BAMM $p_0$ > 0.05" includes Type I error rates for BiSSE analyses conducted only on the set of avian subtrees for which the posterior probability of among-lineage diversification rate variation exceeded 0.95. *N* refers to the total number of simulations, not the number of subtrees; 10 simulations were performed per subtree.

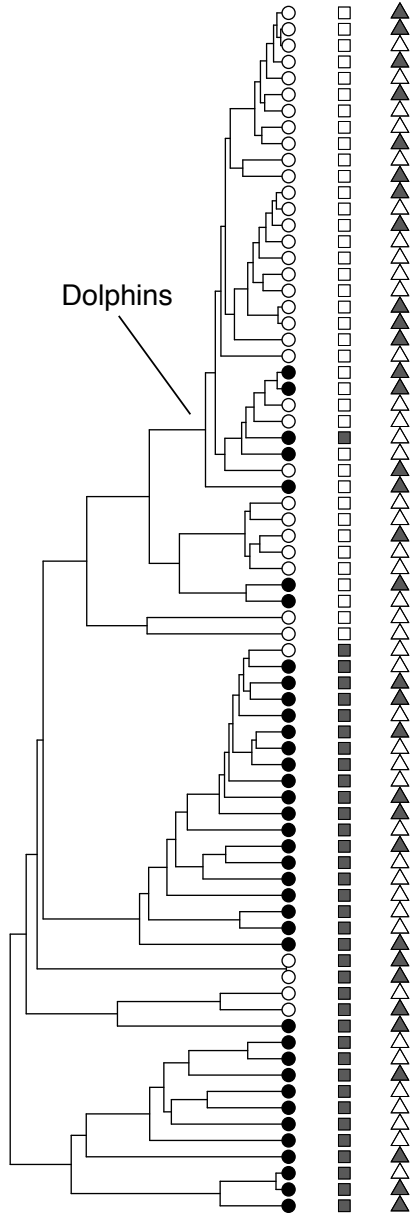

Figure 1. Phylogeny of extant whales from Steeman et al. (2009), showing the observed distribution of small (open circles) and large (filled circles) body size. Squares show a representative distribution of character states simulated under a symmetric Markov model with rare character transitions (q = 0.01; the tree was scaled to a root depth of 1.0), and triangles show a representative distribution simulated under a model with common (q = 10) transitions. The former are clearly co-distributed with body size, and the latter are not.

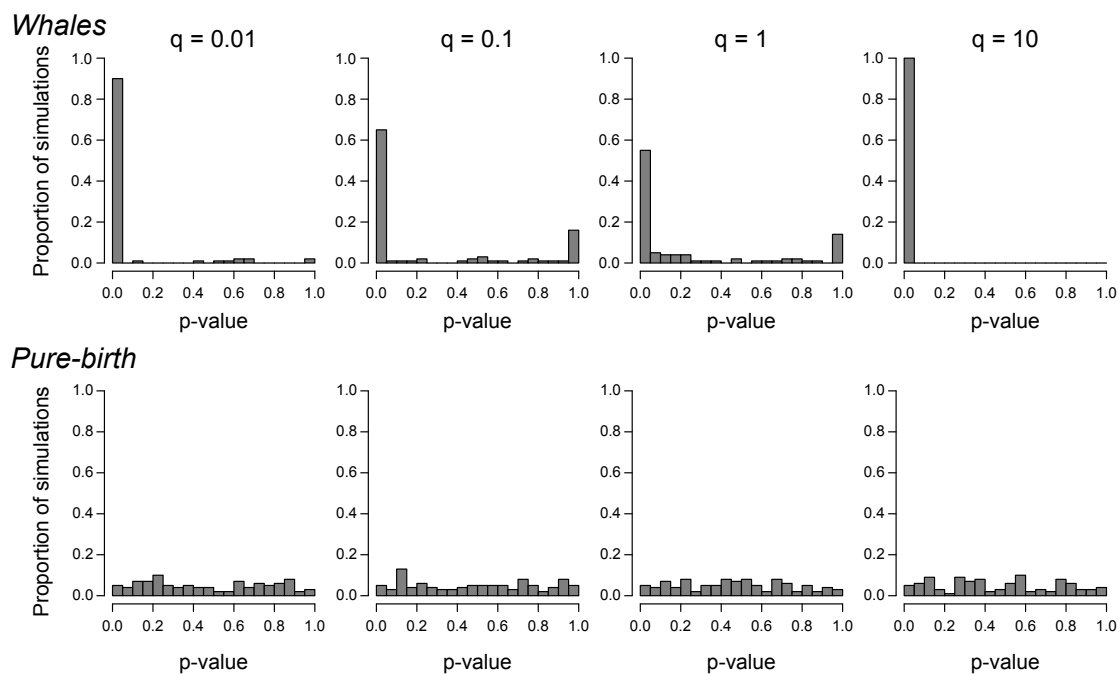

Figure 2. Distribution of p-values for likelihood ratio tests of trait-dependent speciation for character states simulated on the empirical cetacean phylogeny (Fig. 1) (upper row) and for phylogenies of the same size simulated under a pure-birth (PB) process (lower row). All phylogenies were scaled to a root depth of 1.0. Binary characters were simulated on fixed topologies in the absence of trait-dependent speciation or extinction, using a symmetric model with transition rate q. The horizontal axis (p-value) refers to the probability of the data under the null hypothesis that character states have identical speciation rates. For the cetacean phylogeny, the overwhelming majority of simulations incorrectly supported trait-dependent speciation. For the pure-birth phylogenies, with no among-lineage rate variation, error rates are not appreciably elevated.

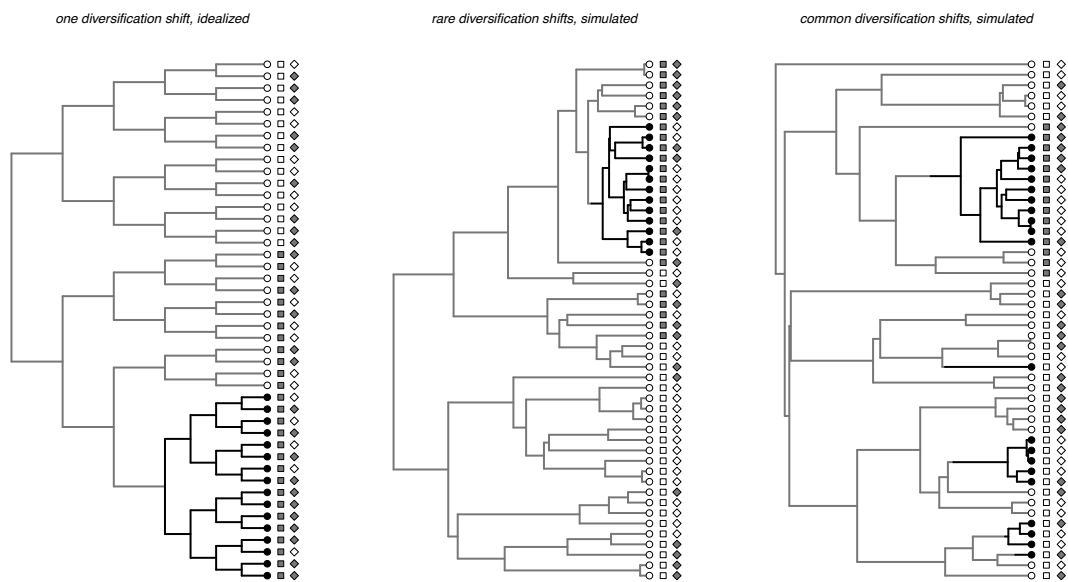

one diversification shift, idealized     rare diversification shifts, simulated     common diversification shifts, simulated

Figure 3. (A) An idealized phylogeny with a single shift in diversification rate. One subclade (black branches) has a speciation rate twice as high as the rest of the clade (gray branches). Symbols at the tips show the state of a character that entirely drives the speciation rate difference (circles). A second character is fixed in the primary clades descended from the root (squares) and is largely co-distributed with the causal trait. A third character is simulated under a model with moderate rates of forward and backward transitions ($q = 1$; diamonds) and is not obviously co-distributed with the others. (B) A simulation analogue of (A). Diversification shifts are driven by one character (circles), which changes state only very rarely. Simulation parameters are $\lambda_0 = 0.5$, $\lambda_1 = 1$, $\mu_0 = \mu_1 = 0$, $q = 0.001$. Two other characters are then simulated on the tree, with either low (squares, $q = 0.01$) or high (diamonds, $q = 1$) transition rates and no influence on speciation. (C) Similar to (B), but with more rapid evolution of the character influencing speciation ($q = 0.1$), and hence more common diversification shifts.

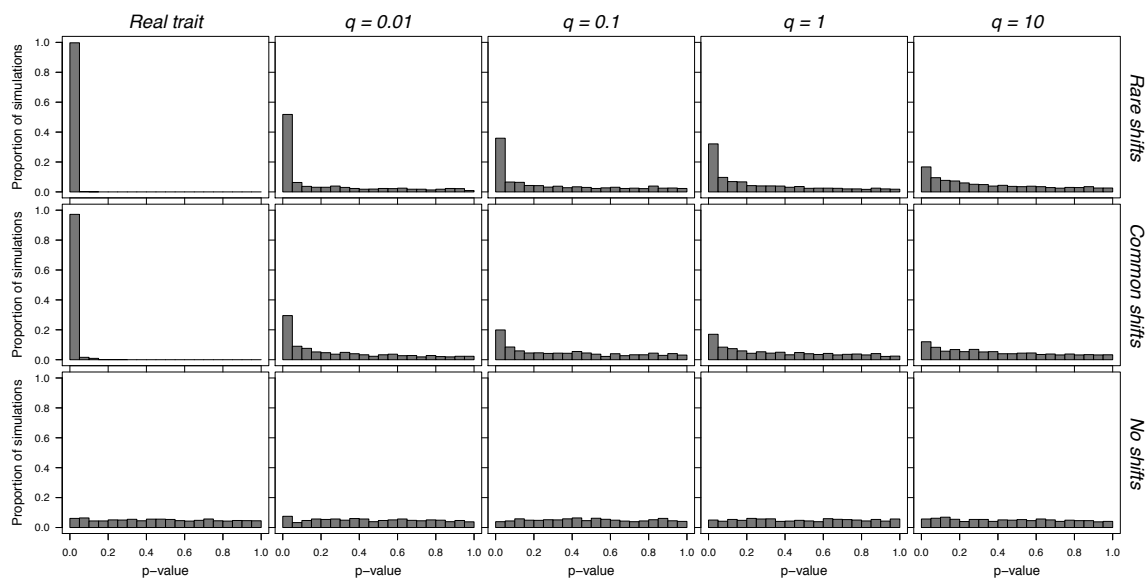

Figure 4. Significance tests for state-dependent speciation, conducted on neutral characters simulated on trees with speciation rate shifts. Trees in the top two rows were generated with the procedure described in Fig 3B and C, respectively, but were more than double the size in those illustrations. Shown are histograms of p-values from a likelihood ratio test of a model with state-dependent speciation ($\lambda_0$, $\lambda_1$, $\mu$, $q_{01}$, $q_{10}$) against a model without ($\lambda_0 = \lambda_1$, $\mu$, $q_{01}$, $q_{10}$). Each panel summarizes results from 1000 trees, each with 200 tips and at least 10% of each character state, and scaled to a root age of 1. All panels in the first row use the same set of trees, on which shifts in diversification rate are rare (simulated with a slowly evolving character influencing speciation, Fig 3B). All panels in the second row use a different set of trees, on which shifts in diversification rate are common (cf. Fig 3C). The first column shows analysis of the trait that truly affects speciation, for which the equal-speciation model is consistently and correctly rejected. The subsequent columns show analyses of neutral characters, simulated with the transition rate shown, q. When the neutral character evolves slowly (q = 0.01 or 0.1), the statistical test frequently but incorrectly concludes the trait is associated with speciation rate differences (Type I error rate of 18-45%). There are many fewer false positives when the neutral trait evolves more rapidly and when shifts in diversification are common on the tree, because these processes help to decouple the neutral trait from the causal one. The third row shows results for a "control" set of simulations, in which there are no diversification shifts on the trees. The distribution of p-values here is approximately uniform, as expected.

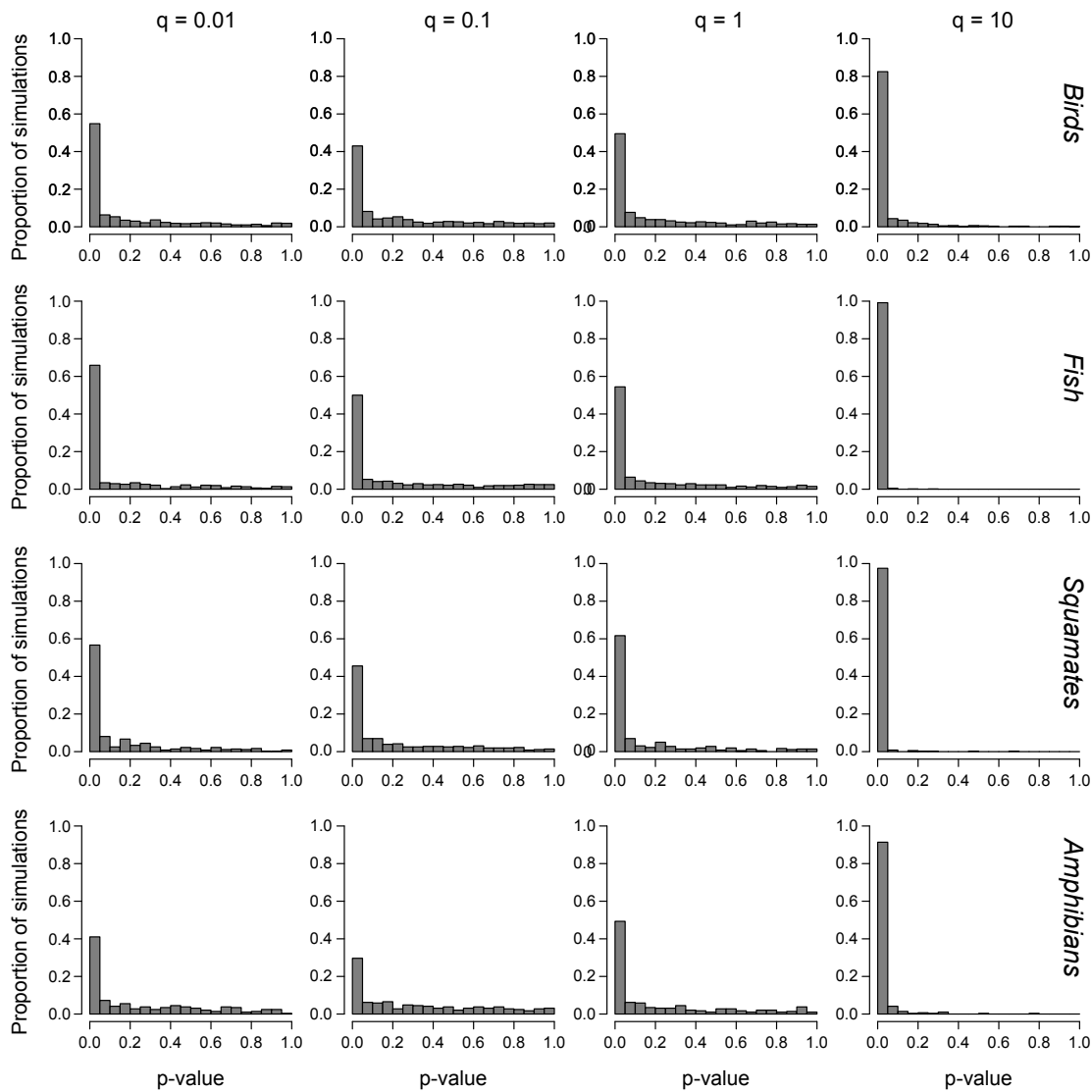

Figure 5. Significance tests for trait-dependent speciation, conducted on neutral characters simulated on subtrees (each with 200 - 500 tips) drawn from the large published phylogenies of birds, fishes, squamates, and amphibians; total numbers of subtrees per taxon were 60 bird trees, 61 fish trees, 36 squamate trees, and 29 amphibian trees. Each panel includes ten simulations per subtree, showing p-values from likelihood ratio tests of the null hypothesis that speciation rates do not differ between character states. All subtrees were scaled to a root depth of 1.0 time units prior to simulation of character states. Results indicate a consistent bias in favor of trait-dependent speciation, even though characters were uncorrelated with speciation rates.

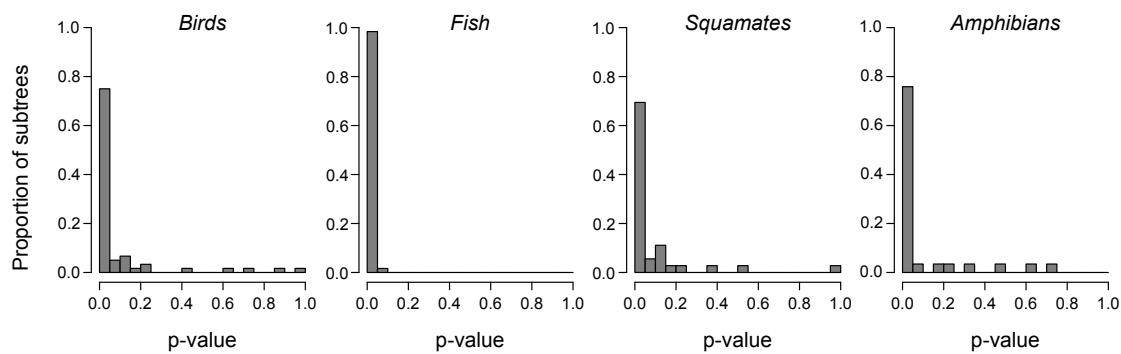

Figure 6. Distribution of p-values for tests of the effect of taxon name length on speciation rate for phylogenies of birds, fishes, squamates, and amphibians; p-value is the probability of the data under the null hypothesis that taxon name length is not associated with speciation rate. Taxon name length was scored as a binary character (short, long) depending on whether the number of letters in the Latin binomial exceeded the median name length in the tree. Phylogenetic subtrees are the same as those used in Fig. 5. Overall, taxon name length is frequently associated with speciation rate.

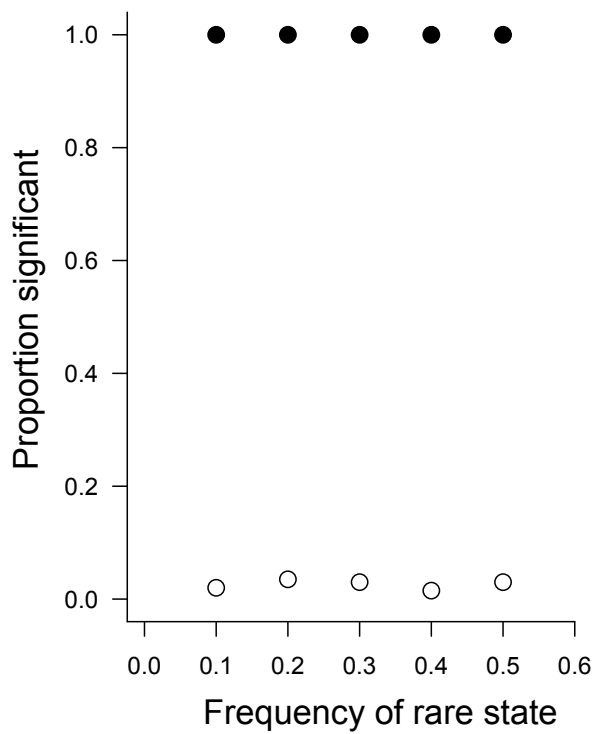

Figure 7. Proportion of simulated datasets showing significant associations between trait and speciation rate (at p < 0.05) for the whale phylogeny (filled circles) and pure-birth phylogenies (open-circles) when character states are unstructured with respect to phylogeny. Simple permutations of traits across the tips of the whale phylogeny results in consistent and strong evidence for trait-dependent diversification.